\documentstyle[prd,aps,preprint]{revtex}
\begin{document}
\tightenlines
\draft
\widetext
\title{Family of  solutions for axisymmetric electrovacuum Einstein-Maxwell
field equations}
\author{J. Socorro{\footnote{E-mail: socorro@ifug3.ugto.mx}} and 
Octavio Cornejo
{\footnote{E-mail: cone@ifug3.ugto.mx}}\\
Instituto de F\'{\i}sica de la Universidad de Guanajuato,\\
Apartado Postal E-143, C.P. 37150, Le\'on, Guanajuato, Mexico.
}

\date{\today}
\maketitle

\begin{abstract}
We present  a family of solutions for the axisymmetric Plebanski-Demianski 
metric and other corresponding reduced metrics. We also present the black hole 
characteristics using a new set of parameters for Kerr-Newman metric.
\end{abstract}
\vspace{0.5cm}

\section{Introduction}
In the last four decades there has been an extraordinary progress in solving
the Einstein-Maxwell equations with two Killing vectors 
\cite{PD,A,AM,FH,H,GM}. 
The most important solutions have been related to charged black holes, i.e., 
Reissner-Nordstr\"om static black hole  and Kerr-Newman rotating 
black hole. In addition, it is well known the solution of 
Plebanski-Demianski (PD) \cite{PD}  metric, containing the last solutions
as particular cases; the PD solution  
is the most general solution in the Petrov classification type D. Besides, 
recently, these solutions were generalized in the metric affine gravity 
\cite{ACMS,AMS,MS}.

The present work is the result of an investigation that we did using
computer algebra methods of the family of solutions of Einstein-Maxwell 
equations with cosmological constant $\lambda$
\begin{equation}
G_{\mu\nu} + \lambda g_{\mu\nu} -\kappa T_{\mu\nu }=0.
\label{fe}
\end{equation}
In particular, we have extensively used REDUCE 3.6 \cite{REDUCE}.

The Maxwell equations can be written in tensorial form as

\begin{eqnarray}
\partial_{\beta}F^{\alpha\beta} &=& j^\alpha, \\
\label{mt1}
\partial_\alpha F_{\beta\theta} + \partial_{\theta} F_{\alpha\beta} 
+ \partial_\beta F_{\theta\alpha} &=& 0,
\label{mt1a}
\end{eqnarray}
where $F^{\alpha\beta}$ is the Maxwell tensor 
\begin{equation}
F_{ab} = \partial_b A_a - \partial_a A_b,
\label{mt3}
\end{equation}
with the quadripotential $A^\alpha$, and  $j^\alpha$  the current
density quadrivector (in this work we take $j^\alpha=0$).

\section{Structure of the PD metric} 
Following  reference \cite{PD}, we consider the class of space-times with 
signature  $(-,+,+,+)$ in real generalized coordinates  
$\chi^\alpha(\tau ,q,p,\sigma )$ with the 
line element given by

\begin{equation}
ds^2= g_{00} (\chi^\alpha)d\tau^2 + 2g_{03} (\chi^\alpha)d\tau d\sigma +g_{11} 
(\chi^\alpha ) dq^2 + g_{22} (\chi^\alpha) dp^2 + g_{33} (\chi^\alpha) 
d\sigma^2, \label{chi}
\end{equation}
where we take only the following two-coordinate dependence of 
$\chi^\alpha = \chi^\alpha (q,p)$. The PD metric has
a similar structure \cite{PD}
\begin{equation}
ds^2 =\frac{1}{H^2} \left\{ -\frac{{\cal Q}}{\Delta} (d\tau -p^2d\sigma)^2 + 
\frac{\Delta}{{\cal Q}} dq^2 +\frac{\Delta}{{\cal P}} dp^2 
+\frac{{\cal P}}{\Delta} (d\tau +q^2 d\sigma )^2 \right\}, \label{ds}
\end{equation}
or
\begin{equation}
ds^2=\frac{{\cal P}-{\cal Q}}{H^2\Delta} d\tau^2 
+ \frac{2(p^2 {\cal Q} +q^2 {\cal P})}{H^2\Delta} d\tau d\sigma 
+\frac{\Delta}{H^2{\cal Q}} dq^2 +\frac{\Delta}{H^2{\cal P}} dp^2 +
\frac{q^4{\cal P}-p^4{\cal Q}}{H^2\Delta} d\sigma^2 . \label{d2}
\end{equation}

The structure functions  $\cal P$ and $\cal Q$ depend on the coordinates 
$\rm p$ and  $\rm q$, respectively. The functions $\rm H$ and $\Delta$ are
defined as $\rm H=1-pq$ and $\rm \Delta = p^2 + q^2$, respectively. This 
space-time has two Killing vectors $\partial /\partial\tau$ and 
$\partial /\partial\sigma$ that commute between them. The vector 
$\partial /\partial\tau$ is timelike in the region where
${\cal Q}-{\cal P}>0$, and the space-time is stationary there.

The real electromagnetic potential is given by
\begin{equation}
\rm A^\alpha=\frac{1}{\Delta} \left( ( e_0 q+ g_0 p),0,0, (g_0 p- e_0p )
pq \right), \label{fra}
\end{equation}
where  $\rm e_0$ and $\rm g_0$ are the electric- and magnetic-like charges, 
respectively.

It is well known that the structure functions $\cal P$ and $\cal Q$ are
polynomials of fourth degree, whose
coefficients, depending on seven parameters, are classified as
mass $\rm m$, NUT parameter $\rm n$, angular momentum
 $\rm j_0$, acceleration parameter $\rm b$,  cosmological constant
 $\lambda$, and the charges $\rm e_0, g_0$.
\begin{eqnarray}
{\cal P} &=& {\rm (b- g^2_0) +2np -\epsilon p^2 +2mp^3 - \left(b+e^2_0 + 
\frac{\lambda}{3}
\right) p^4}, \label{pp} \\
{\cal Q} &=& {\rm (b+ e^2_0) -2mq + \epsilon q^2-2nq^3 - \left(b-g^2_0 
+\frac{\lambda}{3} \right) q^4}, \label{qq} 
\end{eqnarray}
where  $b\to \gamma-\frac{\lambda}{6}$. 

We mention that recently the following generalization of  this metric has 
been found by Garc\'{\i}a and Mac\'{\i}as \cite{AM},
including  a new parameter $\mu$ (new acceleration parameter)

\begin{eqnarray}
{\cal P} &=&  {\rm (b- g^2_0) +2np-\epsilon p^2+2m\mu p^3- 
\left[ \mu^2 (b+e^2_0)+
\frac{\lambda}{3} \right] p^4}, \label{bb} \\
{\cal Q} &=& {\rm (b+e^2_0)-2mq+\epsilon q^2 -2n\mu q^3 - 
\left[ \mu^2 (b-g^2_0)+
\frac{\lambda}{3} \right] q^4}, \label{q1} \\
{\rm H }&=& {\rm  1-\mu pq },\label{h1} \\
{\rm \Delta} &=&{\rm  p^2+ q^2}. \label{dd}
\end{eqnarray}

\section{Family of solution for PD metric}
We present now the family of solutions for Einstein-Maxwell equations with 
cosmological constant $\lambda$

\begin{equation}
\rm G_{\mu\nu} + \lambda g_{\mu\nu} -\kappa T_{\mu\nu} = 0. \label{gg}
\end{equation}

Using the program for Excalc package\cite{exc} in REDUCE 3.6, we can 
introduce the
following ansatz for the polynomials ${\cal P}(\rm p)$ and  
${\cal Q}(\rm q)$
\begin{eqnarray}
{\cal P} (\rm p) &=& {\rm p_0 + p_1 p+ p_2 p^2 + p_3 p^3 + p_4 p^4},
 \label{p0} \\
{\cal Q} (\rm q) &=& {\rm q_0 + q_1 q+ q_2q^2+q_3q^3 + q_4q^4}. \label{q0} 
\end{eqnarray}

Via computer algebra we obtained the following relations

\begin{eqnarray}
{\rm q_0 -p_0} &=& {\rm  e^2_0 + g^2_0}, \label{qp} \\
{\rm p_3}&=& {\rm -\mu q_1},\\
{\rm q_3}&=&{\rm -\mu p_1}, \\
{\rm p_2}&=& {\rm -q_2},  \\
{\rm p_4}&=&{\rm -\frac{\lambda}{3} -\mu^2 q_0}, \\
{\rm q_4}&=&{\rm p_4+\mu^2 (e_0^2+g_0^2)}.
\end{eqnarray}       
At this point we generalize the relation (\ref{qp}) by introducing 
{\it two new real parameters}, $\alpha$ and 
$\beta$, as follows 
\begin{equation}
{\rm q_0 - p_0} = {\rm \left[b+ (\alpha +1) e^2_0 +\beta g^2_0 \right] -
\left[b+ \alpha e^2_0 + (\beta -1) g^2_0 \right] = e^2_0 
+ g^2_0 }. \label{q-0} 
\end{equation}

Finally, the family of solutions given via the structure functions
$\cal P$ and $\cal Q$, can be written by taking into account the
structure function for the PD metric

\begin{eqnarray}
{\cal P} (\rm p)&=&{\rm \left[ b+\alpha e^2_0+(\beta -1)g^2_0\right]+2np
-\epsilon p^2+2m\mu p^3 }\nonumber\\
&& {\rm  - \left[ \mu^2 (b+ (\alpha +1) e^2_0 +\beta g^2_0)+ \frac{\lambda}{3}
\right] p^4} \label{p4}, \\
{\cal Q} (\rm q) &=&{\rm  \left[ b +(\alpha +1) e^2_0 +\beta g^2_0 \right]-2mq 
+\epsilon q^2-2n\mu q^3}\nonumber\\ 
&&{\rm - \left[\mu^2 (b+\alpha e^2_0 + (\beta -1)g^2_0)
+\frac{\lambda}{3} \right] q^4}, \label{pq} \\
\rm  H &=& {\rm 1-\mu pq}, \label{h} \\
\Delta &=& {\rm p^2 + q^2}. \label{d}
\end{eqnarray}
The PD metric corresponds to the following values in the new parameters:
$\mu =1$, $\alpha =0$ and $\beta =0$.

This family of solutions have the property that they remain solutions under 
the inversion transformation 
\begin{equation}
q\rightarrow - \frac{1}{q}
\end{equation}
performed in both the structure functions and the electromagnetic 
potential\cite{PD}. In addition, 
we can find the cylindrically symmetric counterpart of the metric (\ref{ds}) 
in similar form to Garc\'{\i}a et. al, \cite{AMS}. 
%%%%%%%%%%%%%%%%

\section{Roots and singularities in the PD generalized}

When we consider the problem of obtaining the roots and 
singularities in this generalized metric, via calculating the invariants
$\rm R, R_{\alpha\beta}\, R^{\alpha\beta}$, 
$R_{\alpha\beta\theta\delta}\,R^{\alpha\beta\theta\delta}$, we found  
an intrinsic singularity for $\Delta = p^2 + q^2 =0$. Under the coordinate 
transformation
$\rm (q\to r, p\to -{\rm j_0 \cos{\theta}})$, the function $\Delta =0$ gives
the corresponding ring Kerr singularity, where the new parameters
$\alpha$ and $\beta$ do not modify this property.

The roots in the polynomials ${\cal P}=0$ (\ref{p4}) and 
${\cal Q}=0$ (\ref{pq}) represent
singularities in the coordinates, and therefore they are removable via an 
adequate
coordinate transformations. The singularities in the coordinates, show us
some defficiencies in the use of these coordinate systems to
describe the space-time under consideration.

The roots in the polynomial ${\cal P}=0$ do not have any physical interest,
 and  we omit them. However, the real roots in the polynomial ${\cal Q}=0$, 
represent null hypersuperfaces of event horizon type with $\rm q=constant$.

The solution for ${\cal Q}=0$ is obtained via conventional methods
\cite{unspenski}. Consider the following form

\begin{equation}
\left[ b+(\alpha +1) e^2_0 + \beta g^2_0 \right] -2mq+\epsilon q^2
-2n\mu q^3 - \left[\mu^2 (b+\alpha e^2_0 +(\beta -1)g^2_0) +
\frac{\lambda}{3} \right] q^4 = 0, \label{q=} 
\end{equation}
or,
\begin{equation}
q^4 + a_3 q^3 + a_2 q^2 + a_1 q+ a_0 =0, \label{q+}
\end{equation}
where

\begin{eqnarray}
a_3 &=& \frac{2n\mu}{\mu^2 (b+\alpha e^2_0 + (\beta -1) g^2_0)+
\frac{\lambda}{3} }, \nonumber \\
a_2 &=& - \frac{\epsilon}{ \mu^2 (b+\alpha e^2_0 + (\beta -1) g^2_0)+
\frac{\lambda}{3} }, \nonumber \\
a_1 &=& \frac{2m}{ \mu^2 (b+\alpha e^2_0 + (\beta -1) g^2_0)+
\frac{\lambda}{3} }, \nonumber \\
a_0 &=& - \frac{[b+ (\alpha + 1) e^2_0 +\beta g^2_0}{ \mu^2 (b
+ \alpha e^2_0 + (\beta -1) g^2_0)+
\frac{\lambda}{3} } \, . \nonumber 
\end{eqnarray}
Using the Ferrari method for equations of fourth degree and the Cardano method
for  those of third degree \cite{unspenski}, we can get the following roots 
\begin{equation}
q_{a ,b} =\frac{1}{2} (-A\pm B^{1/2}), \qquad 
q_{c ,d} = \frac{1}{2} (-C\pm D^{1/2}), \label{qga}
\end{equation}
where
\begin{eqnarray}
A &=& \left[ \frac{a_3}{2}- \frac{a^2_3}{A}+a_2 - (t_1)^{1/3} - (t_2)^{1/3}
\right] ,\nonumber \\
B &=& \left\{ A^2-2 \left\{ (t_1)^{1/3} + (t_2)^{1/3} -
\frac{\frac{1}{2} a_3 [(t_1)^{1/3} + (t_2)^{1/3}]-a_1}{\left(\frac{a_3}{2}
-A\right)^{1/2}} \right\}\right\}, \nonumber \\
C &=& \left[ \frac{a_3}{2} + \frac{s^2_3}{4} -a_2 + (t_1)^{1/3} + (t_2)^{1/3}
 \right], \nonumber \\
D &=& \left\{ C^2-2 \left\{ (t_1)^{1/3} + (t_2)^{1/3} +
\frac{\frac{1}{2} a_3 [(t_1)^{1/3} + (t_2)^{1/3}]-a_1}{\left(\frac{a_3}{2}
-A\right)^{1/2}} \right\}\right\}, \nonumber \\
t_1 &=& - \frac{u}{2} + \left(\frac{u^2}{4} + \frac{\upsilon^3}{27} 
\right)^{1/2} , \quad t_2 = u - t_1, \nonumber \\
u &=& b_1 - \frac{b^2_2}{3} , \qquad \upsilon = b_0 -\frac{b_1 b_2}{3}
+ \frac{2b^2_2}{27} ,\nonumber \\
b_0 &=& \frac{4}{q^3_4} \left\{ \epsilon q_0 q_4 + n^2 \mu^2 q_0 -4 m^2 q_4 
\right\}, \nonumber \\
b_1 &=& \frac{4}{q^2_4} \left\{ mn\mu + 4 q_0 q_4 \right\} , \quad
b_2 = \frac{\epsilon}{q_4}, \nonumber \\
q_0 &=& \left[ b+ \alpha e^2_0 + (\beta -1) g^2_0 \right], \nonumber \\
q_4 &=& \left[ \mu^2 (b+ (\alpha +1) e^2_0 + \beta g^2_0) 
+ \frac{\lambda}{3} \right]. \nonumber 
\end{eqnarray}

The polynomial  ${\cal Q}$ (\ref {q=}) can be rewritten as
\begin{equation}
 \rm q_4 \left\{ q^4+a_3 q^3 + a_2 q^2+a_1q+a_0 \right\} = q_4 (q-q_I)
(q-q_{II}) (q-q_{III}) (q-q_{IV}), \label{qq4}
\end{equation}
where $\rm q_I, q_{II}, q_{III}$ and $\rm q_{IV}$ are the roots (\ref {qga}) 
in ascendent order of magnitude, satisfying the following
relations 
\begin{eqnarray}
\sum^{IV}_{i=I} q_i &=&-a_3 =- \frac{2n\mu}{\left[\mu^2 \left(b+\alpha e^2_0
+ (\beta -1) g^2_0\right) +\frac{\lambda}{3}\right]}, \label{2n} \\
\sum_{i>j} q_i q_j &=&a_2=- \frac{\epsilon}{\left[\mu^2 (b+\alpha e^2_0 
+(\beta -1) g^2_0)+ \frac{\lambda}{3} \right]}, \label{eps} \\
\sum_{i>j>k} q_i q_j q_k &=&-a_1=- \frac{2m}{\left[\mu^2 (b+\alpha e^2_0 + 
(\beta-1) g^2_0)+ \frac{\lambda}{3} \right]} ,\label{2m} \\ 
\Pi_i q_i=a_0&=&-\frac{b+(\alpha +1) e^2_0 +\beta g^0_2}{\left[\mu^2 
(b+\alpha e^2_0+(\beta-1) g^2_0)+\frac{\lambda}{3} \right]}, \label{b+a}
\end{eqnarray}
where $\rm a_i, i=0,1,2,3$ are the coefficients of ${\cal Q}$.

Also, we can obtain red-shifted hypersuperfaces for this generalized
solution, which are given by  the equation \cite{Wald}
\begin{equation}
g_{00} = \frac{P-Q}{H^2\Delta} = 0.
\label{ro}
\end{equation}
The roots of (\ref{ro}) are analogue in form to those in  equations 
(\ref {qga}), except that the coefficient $a_0$ is now a function of the 
coordinate p.

\section{Reduced family of solutions from the family of solutions to PD 
metric}
Certain family of reduced solutions can be obtained from the family of 
solutions of the PD metric when we drop appropriately the new parameters
 and perform the coordinate transformations from 
$\rm (\tau,p,q,\sigma) \to $ (other coordinates) as used in Garc\'{\i}a 
\cite{A}. 
We give these reduced generalized metrics
in table I. All of them correspond to stationary axisymmetric space-times
with Maxwell field as a matter source.

\begin{center}
\begin{tabular}{|c|}\hline
Metric \\
$ ds^2 = \frac{1}{H^2} \left\{ -\frac{{\cal Q}}{\Delta} (d\tau 
-p^2 d\sigma )^2 + \frac{\Delta}{{\cal Q}} dq^2 
+ \frac{\Delta}{{\cal P}} dp^2 + \frac{{\cal P}}{\Delta} (d\tau +
q^2 d\sigma )^2 \right\} $\\ \hline
 electromagnetic potential \\
$ A=\frac{1}{\Delta} \left((e_0 q+ g_0 p), 0,0, (g_0p- e_0 q)pq\right)$\\ 
\hline
\end{tabular}
%\end{center}
%\begin{center}
\begin{tabular}{|c|c|}\hline 
Metric & structure functions \\ \hline
PD generalized &  ${\cal P}=\left[b+\alpha e^2_0+ (\beta -1)g^2_0\right] 
+2np - \epsilon p^2+2m\mu p^3  $\\
(10 parameters)  &$-\left[\mu^2(b +(\alpha +1)e^2_0  + \beta g^2_0)
+\frac{\lambda}{3}\right] p^4 $\\
 & $ {\cal Q}=\left[ b+(\alpha +1)e^2_0+ \beta g^2_0 \right]
-2mq+ \epsilon q^2-2n\mu q^3$ \\
 & $\qquad\quad -\left[\mu^2(b+\alpha e^2_0+(\beta -1)g^2_0)+
\frac{\lambda}{3} \right] q^4 $  \\
 & $ \Delta = p^2 + q^2$  \\
 & $H= 1-\mu pq$ \\ \hline
PD standard & ${\cal P}=(b-g^2_0)+2np-\epsilon p^2 +2mp^3- 
(b+e^2_0 + \frac{\lambda}{3}) p^4$ \\ 
$(\mu =1,\alpha =0,\beta =0)$ & ${\cal Q}=(b+e^2_0)-2mq +\epsilon q^2-2nq^3-
(b- g^2_0 +\frac{\lambda}{3}) q^4$ \\
& $\Delta = p^2 + q^2$ \\
& $H= 1-pq$ \\ \hline
PC generalized & ${\cal P}= \left[b+\alpha e^2_0 +(\beta -1)g^2_0\right] 
+ 2np-\epsilon p^2 - \frac{\lambda}{3} p^4$ \\
$(\mu =0)$ & ${\cal Q}=\left[b+(\alpha +1) e^2_0 +\beta g^2_0\right] -2mq
+\epsilon q^2 - \frac{\lambda}{3} q^4$ \\
& $\Delta = p^2 + q^2 $ \\
& $H=1$ \\ \hline
KN generalized & ${\cal P}=\left[j_0^2+\alpha e^2_0+ (\beta-1)g^2_0\right]
-(1-\frac{\lambda j_0^2}{3}) p^2 -\frac{\lambda}{3} p^4$ \\
$(\mu =0,n=0,\lambda\neq 0$ & ${\cal Q}=\left[j_0^2+(\alpha +1)e^2_0+\beta g^2_0
\right]-2mq+ (1-\frac{\lambda j_0^2}{3}) q^2 -\frac{\lambda}{3} q^4$ \\
$b\to j_0^2, \epsilon \to (1-\frac{\lambda j_0^2}{3}))$ & $\Delta = p^2 
+ q^2$\\
& $H=1$ \\ \hline
KN generalized & ${\cal P}=\left[j_0^2+\alpha e^2_0 + (\beta -1) g^2_0\right]
-p^2$ \\
$(\mu =0,n=0,\lambda =0,$ & ${\cal Q}=\left[j_0^2+ (\alpha +1) e^2_0+\beta g^2_0
\right] -2mq+ q^2$ \\
$b\to j_0^2, \epsilon \to1)$ & $\Delta = p^2 +q^2$
 \\
& $H=1$ \\ \hline
KN standard & ${\cal P}= j_0^2 - p^2$ \\
$\mu =0,n=0, \lambda =0,$ & ${\cal Q}= (j_0^2+ e^2_0 +g^2_0)-2mq+q^2$ \\
$\alpha =0, \beta =1$, & $\Delta = p^2 + q^2$ \\
$b\to j_0^2, \epsilon\to 1$ & $H=1$ 
\\ \hline
\end{tabular} 
\end{center}
\begin{center}
{\small Table I\\
Generalization and standard cases contained in the PD solutions}
\end{center}

%%%%%%%%%%%%%%%%%%%%%%%%%%%%%

\section{What to do the new parameters  $\alpha$ and $\beta$? }
In order to study the features of the new parameters $\alpha$ and $\beta$, we
take the KN metric as an example. As it is known, the exterior gravitational 
field of a black hole bounded by an event horizon is 
characterized by three
parameters: mass m, angular momentum $\rm j_0$, and the electric charge 
$\rm e_0$.

When we take the Boyer-Lindquist coordinate transformation  in the 
generalized PD metric, we obtain the generalized KN metric (with $\lambda=0$)
through the transformation

\begin{equation}
T:\left\{\tau\to t-j_0 \phi ,q\to r,p\to - j_0 \cos \theta ,
\sigma\to -\phi/j_0\right\}, 
\end{equation}
\noindent
where now the space-time have the following coordinate 
$\chi^\mu =(t,r,\theta ,\phi)$, and the explicit form of the metric is

\begin{eqnarray}
ds^2 &=& \frac{{\cal P}_{KN}-{\cal Q}_{KN}}{\Delta_{KN}}dt^2
-\frac{2}{a\Delta_{KN}}\left\{(j_0^2+r^2){\cal P}_{KN} 
-j_0^2 \sin^2 \theta {\cal Q}_{KN}\right\} dtd\phi 
+  \frac{\Delta_{KN}}{{\cal Q}_{KN}} dr^2 \nonumber \\
&+& \frac{j_0^2 \sin^2\theta \Delta_{KN}}{{\cal P}_{KN}} d\theta^2 
+ \frac{1}{j_0^2\Delta_{KN}} \left\{ (j_0^2+r^2)^2 {\cal P}_{KN} 
-j_0^4 \sin^4 \theta {\cal Q}_{KN} \right\}d\phi^2 ,
\label{metkn}
\end{eqnarray}
with the structure functions ${\cal P}_{KN}$, ${\cal Q}_{KN}$ and 
$\Delta_{KN}$ given by
\begin{eqnarray}
{\cal P}_{KN} &=& \left[ \alpha e^2_0+(\beta 
-1)g^2_0 \right]+j_0^2 \sin^2\theta , \\ 
{\cal Q}_{KN} &=& \left[ j_0^2+(\alpha +1)e^2_0 +\beta g^2_0 \right]-2mr+r^2,
 \\
\Delta_{KN} &=& r^2 + j_0^2 \cos^2 \theta. 
\end{eqnarray}

The hypersurfaces of infinite shift would satisfy the condition
\begin{equation}
g_{tt} = \frac{{\cal P}_{KN} - {\cal Q}_{KN}}{\Delta_{KN}} = 0,
\end{equation}
impliying that there exist two  such  hypersurfaces at the radial distance 
\begin{equation}
r=m\pm \left[ m^2- (e^2_0 +g^2_0+j_0^2\cos^2\theta )\right]^{1/2},
\label{r1}
\end{equation}
where the equation
\begin{equation}
0<  e^2_0 +g^2_0 +j_0^2 \leq m^2,
\label{erg1}
\end{equation}
gives us the condition for non existence of imaginary values and of negative 
radius.

Examining the quadratic invariants, we find
an intrinsic ring singularity for $\Delta_{KN}=0$ that is similar to Kerr's
singularity

\begin{equation}
\Delta_{KN} = r^2 +j_0^2 \cos^2\theta =0.
\end{equation} 
\noindent
As is usual, the event horizon can be calculated from the equation
${\cal Q}_{KN} =0$, leading to
\begin{equation}
r=m\pm \left[m^2- ((\alpha +1) e^2_0+\beta g^2_0+j_0^2)\right]^{1/2}.
\label{rkn}
\end{equation}
\noindent
We can conclude that the parameters $\alpha$ and $\beta$ modify the radius 
of the event horizon of the generalized solution of
KN metric. We investigate this  modification in the following.

\subsection{Parametrization of the event horizon in the KN space time}

The ideal case under consideration corresponds to the equation  (\ref {rkn}), 
where there are no event horizon with imaginary or negative radius. Therefore, we have the condition
\begin{equation}
0< (\alpha +1) e^2_0 + \beta g^2_0 + j_0^2 \leq m^2
\end{equation} 
or

\begin{equation}
- (e^2_0 + j_0^2) <\alpha e^2_0 +\beta g^2_0 \leq m^2 - (e^2_0 + j_0^2). 
\label{alfa1}
\end{equation}

%%%%%%%%%%%%%%%%%%
\vglue -1cm 
 \setlength{\unitlength}{0.01cm}
\begin{picture}(1000,400)(200,300)
\thicklines
\put (750,200){\line(1, 0){580}}
\put (300,200){\line(1, 0){580}}
\put (750,200){\line(0, 1){350}}
\put (750,0){\line(0, 1){350}}
\put (1250,170){\line(-2,1 ){650}}
\put (1150,130){\line(-2,1 ){600}}
\put (1000,0){\line(-2,1 ){650}}
\put (750,600){\makebox(0,0){${\rm e_0^2 \, \alpha}$}}
\put (1420,200){\makebox(0,0){${\rm g_0^2 \, \beta}$}}
\put (780,450){\makebox(0,0){${\rm P_1}$}}
\put (750,420){\makebox(0,0){$\bullet$}}
\put (780,350){\makebox(0,0){${\rm P_2}$}}
\put (750,330){\makebox(0,0){$\bullet$}}
\put (780,130){\makebox(0,0){${\rm P_4}$}}
\put (750,120){\makebox(0,0){$\bullet$}}
\put (1020,230){\makebox(0,0){${\rm P_3}$}}
\put (1010,200){\makebox(0,0){$\bullet$}}
\put (1200,550){\makebox(0,0){$\rm P_1(0,m^2-(e_0^2+j_0^2))$}}
\put (1090,500){\makebox(0,0){$\rm P_2(0,g_0^2)$}}
\put (1080,450){\makebox(0,0){$\rm P_3(g_0^2,0)$}}
\put (1160,400){\makebox(0,0){$\rm P_4(0,-(e_0^2+j_0^2))$}}
\put (1250,-20){\makebox(0,0){$\rm e_0^2\,\alpha + g_0^2 \, 
\beta=-(e_0^2+j_0^2)$}}
\put (1350,50){\makebox(0,0){$\rm e_0^2\,\alpha + g_0^2 \, \beta < 
m^2 -(e_0^2+j_0^2)$}}
\put (1400,150){\makebox(0,0){$\rm e_0^2\,\alpha + g_0^2 \, \beta=
m^2-(e_0^2+j_0^2)$}}
\end{picture}
\vglue 3.7cm
\begin{center}
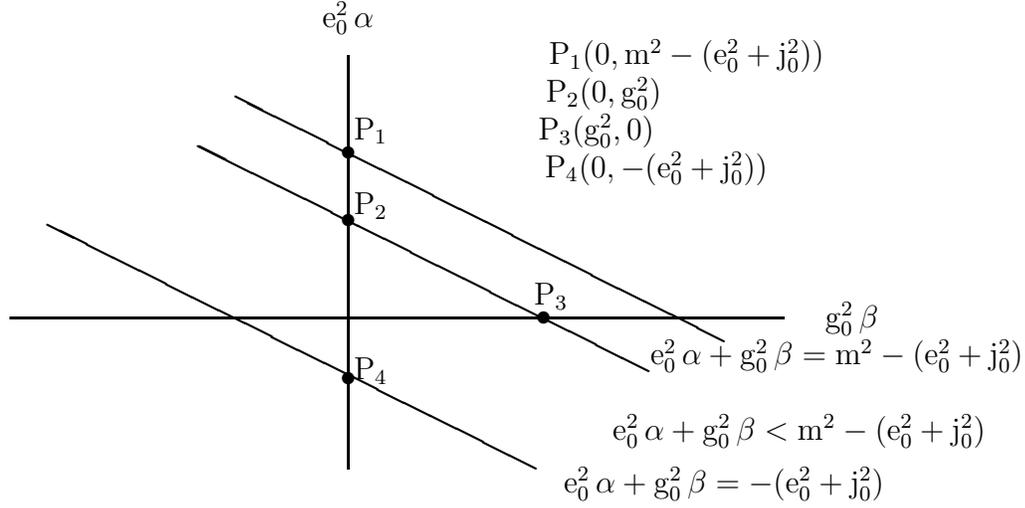
{\footnotesize Fig. 1.: Family of solutions to the equation (\ref {alfa1}).
 }
\end{center}

The set of solutions shown in Figure 1 represents a family of straight lines
of slope $\gamma=-1$. If the values for $\alpha$ and 
$\beta$ given at the points $P_2$ or $P_3$ are introduced in the metric 
(\ref{metkn}), the standard KN solution and the corresponding event horizons 
are recovered. In fact, any value for $\alpha$ and $\beta$ coinciding with a 
straight line crossing through the points $P_2$ and $P_3$ will lead to the 
standard KN metric. Then, the family of parallel straight lines given by
the equation (\ref{alfa1}) represents different KN space-times 
for each case, each with its own event horizon.

Any straight line that crosses the solution range of  $\alpha$ and $\beta$, 
intersecting the family of parallel lines of slope $\gamma$, will lead to 
all spacetimes found in this range
of solutions. Then we  suggest that there is a simplification in
the set of parameters if we take the straight line as the axis
$\alpha \, (\beta=0)$ and introduce a new parameter $\nu$  that reruns this
set of solutions.
 We propose the following form of this straight line 

\begin{equation}
e^2_0 \alpha =m^2\nu - (e^2_0 + j_0^2), \qquad \qquad  0<\nu \leq 1.
\label{nu1}
\end{equation}   
 Thus, the new structure functions are
\begin{eqnarray}
{\cal P}_{KN} &=& \left[m^2 \nu -(e^2_0+g^2_0 + j_0^2)\right] 
+ j_0^2 \sin^2\theta, \\
\label{fenu1}
{\cal Q}_{KN} &=& m^2 \nu -2mr+ r^2, \\
\label{fenu2}
\Delta_{KN} &=& r^2 + j_0^2 \cos^2 \theta,
\end{eqnarray}
\noindent
where  $\nu$ satisfies equation (\ref {nu1}).

All black hole properties  mentioned above are mantained with the
inclusion of this new parameter and the important point is that this
solution, it not violate no-hair theorem.

The equation ${\cal Q}_{KN} = 0$ for the structure function 
(\ref {fenu2}) determines the event horizons whose radiis  are given by 
\begin{equation}
r_{\pm} = m[1 \pm \sqrt{1- \nu}], \qquad\qquad 0< \nu \leq 1 .
\label{rnu1}
\end{equation}
\noindent
Therefore, from the range of $\nu$ we get the following values for the interior
$\rm r_-$ and exterior $\rm r_+$ radiis 
\begin{equation}
0 < r_{-} \leq m , \qquad\qquad m \leq r_{+} < 2m.
\label{rnu3}
\end{equation}
One can see that the corresponding solutions
are proportional to the  geometric mass.
The results given in equations (\ref {rnu1})
-(\ref {rnu3}) represent the parametrization for the event horizon
of KN type black hole, where each value of $\nu$ gives a different
space-time with the interior and exterior horizon. 

\subsection{Quantitave analysis  to the KN space-times and its
event horizons}
It is easy to see that with 
$\nu=(e_0^2 + g_0^2 + j_0^2)/m^2$ we can recover the  standard KN 
space-time. 

Equation (\ref {erg1}) gives us two ideal cases 
\begin{equation}
0<e_0^2 + g_0^2 + j_0^2 < m^2,
\label{erg2}
\end{equation}
and
\begin{equation}
e_0^2 + g_0^2 + j_0^2 = m^2.
\label{erg3}
\end{equation}
The set of solutions for the KN spacetimes taking into account the parameter 
$\nu$, 
 are shown in  figure 2.
%%%%%%%%%%%%%%%%%%
\vglue -.5cm 
 \setlength{\unitlength}{0.01cm}
\begin{picture}(1000,400)(200,300)
\thicklines
\put (500,200){\line(1, 0){800}}
\put (500,500){\line(1, 0){600}}
\put (500,200){\line(0, 1){390}}
\put (800,200){\line(0, 1){300}}
\put (1100,200){\line(0, 1){300}}
\put (1100,350){\line(-4,1 ){600}}
\put (1000,410){\makebox(0,0){${\rm r_+ }$}}
\put (1050,480){\makebox(0,0){${\rm II }$}}
\put (700,480){\makebox(0,0){${\rm I }$}}
\put (1100,350){\line(-4,-1 ){600}}
\put (1000,290){\makebox(0,0){${\rm r_- }$}}
\put (1380,210){\makebox(0,0){${\rm m \nu^2 }$}}
\put (1100,170){\makebox(0,0){${\rm m^2 ( \nu=1) }$}}
\put (800,170){\makebox(0,0){${\rm (e_0^2+g_0^2+j_0^2 ) }$}}
\put (400,510){\makebox(0,0){${\rm m^2 }$}}
\put (370,580){\makebox(0,0){${\rm e_o^2+g_0^2+j_0^2}$}}
\put (330,350){\makebox(0,0){${\rm e_o^2+g_0^2+j_0^2< m^2}$}}
\put (1700,200){\line(0, 1){350}}
\put (1600,560){\makebox(0,0){${\rm r=m\left[1\pm\left(1-\nu 
\right)^{\frac{1}{2}} \right] }$}}
\put (1690,200){\line(1, 0){30}}
\put (1745,200){\makebox(0,0){${0}$}}
\put (1640,275){\makebox(0,0){${r_-}$}}
\put (1690,500){\line(1, 0){30}}
\put (1750,500){\makebox(0,0){${2m}$}}
\put (1640,425){\makebox(0,0){${r_+}$}}
\put (1690,350){\line(1, 0){30}}
\put (1745,350){\makebox(0,0){${m}$}}
\end{picture}
\vglue 1.7cm
\begin{center}
{\footnotesize Fig. 2.: .
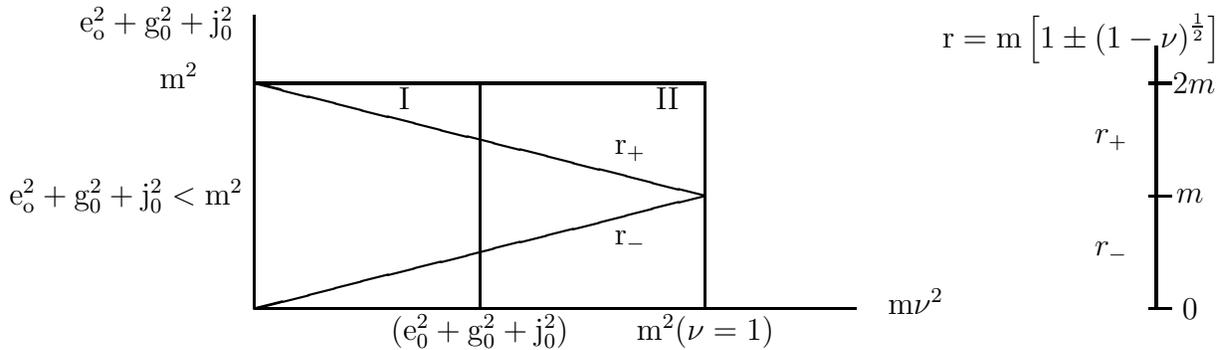
Set of solutions for the KN space-times for the case
 $0 < e_0^2 + g_0^2 + j_0^2 < m^2$. The  superposed straight lines $r_{-}$ 
and $r_{+}$ represent the  radius of the interior and exterior event horizon,
respectively. }
\end{center}
To analyse this figure, we can  divide it in two regions
\begin{eqnarray}
I:&& 0<\nu\leq (e_0^2+g_0^2 + j_0^2)/m^2 \nonumber\\
II:&& (e_0^2+g_0^2 + j_0^2)/m^2 \leq \nu \leq 1.
\end{eqnarray}

The first region I is not physical, because the event horizons cross
 the infinite redshift surface in different points of the rotation axis, 
and the corresponding angles of intersection for both surfaces violate the 
range for the function $\sin^2x$ ($0 \leq \sin^2x \leq 1$).

The region II describes the physical parametrization, because it
contains the family of KN space-times with their event horizons 
of  radiis
\begin{equation}
r_{KN\pm} = m[1 \pm \sqrt{1 - \nu}].
\label{rnu4}
\end{equation}
The set of spacetimes corresponding to the value $\nu = 1$, represents 
non-extreme black holes with only one event horizon at $r_{KN} = m$.

\section{Conclusions}
%%%%%%%%%%%%%%%%%%%%%%%%%%%%%%%%

	Although, we dealt with a difficult algebra we have  been able to 
obtain a generalization of the PD metric by extensive use of the REDUCE 3.6 
system. The latter could be used in any available computer resources.

We introduced a PD metric structure with 10 parameters and 
an appropriate reduction of parameters in the generalized PD metric leads to
 obtain two new families of solutions to the Plebanski-Carter and
Kerr-Newman metrics, containing the new parameters $\alpha$ and $\beta$.
We point out that the introduction of the parameters $\alpha$ and 
$\beta$ does not modify the generalized solutions when the electromagnetic 
charge parameters $e_0$ and $g_0$ are interchanged, as it does in the 
structure functions and the electromagnetic potential considering all their
posible combinations.

	Interpretation of the generalized KN metric as the exterior field of
a collapsed star surrounded by an event horizon, i.e., as a black hole, 
has lead to a new equivalent generalized metric, but now simplified with  only 
one dimensionless parameter called $\nu$. In addition, we have determined 
the way $\nu$ modifies the radius of event horizon such that the 
corresponding solutions
are proportional to the  geometric mass. On the other hand, the redshift 
surfaces remain constant and independent of $\nu$.

	The minimum value of the parameter $\nu$ corresponds to the 
space-time of a standard KN type black hole. Increments in the value of 
$\nu$ increase the 
region between the redshift surfaces and the event horizons (the ergospheres)
until the maximum value for $\nu=1$. This maximum value corresponds to 
the space-time of a KN type black hole with  only one event horizon which is 
not necessarely an extreme black hole. The space-time of a standard extreme KN 
type black hole can also be recovered.

 Dropping the electromagnetic charge parameters 
allows to get a generalized Kerr solution that corresponds to a vacuum 
stationary axisymmetric solution of Einstein equations. The analysis and 
interpretation as a black hole and parametrization of the event horizons is 
completely analogous to the generalized KN case.
%%%%%%%%%%%%%%%%%%%%%%%%%%%%%%%%%%

\noindent {\bf Acknowledgments}\\
We are grateful to Haret Rosu for useful and constructive comments.
This research was supported partially by  CONACyT.

\end{document}